\documentstyle[11pt,aaspp,psfig]{article}
\def\zs{z_{S}}
\def\zl{z_{L}}

\def\ml{m_{L}}
\def\thc{\theta_{crit}}

\begin{document}

\title{\large\bf
 A Measurement of the Cosmological Constant \\
 Using Elliptical Galaxies as Strong Gravitational Lenses}
\author{\bf Myungshin Im, Richard E. Griffiths, \& Kavan U. Ratnatunga \\
  Dept. of Physics \& Astronomy, Johns Hopkins University, \\
  Baltimore, MD 21218}

\begin{abstract}
 We have identified seven (field) elliptical galaxies acting as strong
gravitational lenses and have used them to measure cosmological
parameters.
  To find the most likely value for $\Omega_{m}$ ($=\Omega_{matter}$)
 and $\Lambda$, we have used the combined probabilities of these lens systems
 having the observed critical radii (or image deflection) for the measured
 or estimated values of lens redshifts, source redshifts, and lens
 magnitudes.
  Our measurement gives
$\Lambda=0.64^{+0.15}_{-0.26}$ if $\Omega_{m}+\Lambda=1$, and the
$\Omega_{m}=1$ model is excluded at the 97 \% confidence level.
 We also find, at the 68 \% ($\Omega=0$) -- 82 \% ($\Omega=0.3$)
 confidence level, that an open universe is less likely than a flat
 universe with non-zero $\Lambda$.
 Except for the possibility of strong perturbations due to
 cluster potentials and the systematic overestimate of the lens magnitudes,
 other possible systematic errors do not seem to influence our results
 strongly: correction of possible systematic errors
 seems to increase the significance
 of the result in favor of a non-zero $\Lambda$ model.
\end{abstract}

\keywords{ cosmology: theory - cosmology: observations - cosmology:
 gravitational lensing - galaxies: evolution }

{\it Submitted to Ap.J Feb. 29 1996, Accepted Aug. 20 1996}

\section{Introduction}

Recently, non-zero cosmological constant ($\Lambda$) models have found
increased popularity (e.g., Ostriker \& Steinhardt 1995) owing to the
problem of the age discrepancy implied by the latest Hubble Space Telescope
(HST) measurements of the Hubble constant:
the ages of globular clusters are apparently
larger than the age of the universe predicted by the standard
$\Omega_{m}=1$ model which is favored by standard inflationary
theory (e.g., Freedman et al. 1994; for a brief overview of these arguments,
 see Rees 1996).

In order to measure $\Lambda$, it has been suggested that strong
gravitational lenses might be used, i.e. isolated galaxies or clusters
of galaxies for which the gravitational potential results in multiple
imaging of a background object (Paczynski \& Gorski 1981; Alock \&
Anderson 1986; Gott, Park, \& Lee 1989).  Following these suggestions,
the use of the lens number counts (or the optical depth) was advocated
since this is very sensitive to $\Lambda$ (Turner 1990; Fukugita,
Futamase, \& Kasai 1990; Fukugita et al. 1992).  Maoz \& Rix (1993) and
Kochanek (1995) have applied this method, obtaining upper limits of $\Lambda
\lesssim 0.7$.

Kochanek (1992) has also suggested the lens redshift method, which,
compared with the method based on lens counts,
requires less presumptions about the properties of lenses
and sources, properties which might bias the lens counts considerably
(Helbig \& Kayser 1995; Kochanek 1992; Fukugita \& Peebles 1995).
Taking into account the
selection effects which were neglected in his early study (Kochanek
1992: for a discussion on this selection effect, see Helbig \& Kayser
1995), Kochanek (1995) finds $\Lambda < 0.9$ at $2\,\sigma$ with a peak
at $\Lambda=0.4$.  However, the estimated value of $\Lambda$ in Kochanek
(1995) is sensitive to the detection threshold which is not well understood,
and thus it can not be considered very seriously at the present stage.

It has been recognized that the mean splitting of the lensed images
alone is useful for studies of the dynamical properties of lens
galaxies, but not for the measurement of $\Lambda$ (Turner, Ostriker, \&
Gott, hereafter TOG84; Fukugita et al. 1992).  However, when the mean
separation is used together with other information such as the lens
redshift, the lens magnitude, and the velocity dispersion of the lens
galaxy, then the mean separation does become sensitive to $\Lambda$
(Paczynski \& Gorski 1981; Gott et al. 1989; Kochanek 1992;
Miralda-Escude 1991).  In this {\it Letter}, we will try to measure
$\Lambda$ using a method which we call the ``lens parameter method'',
which is basically similar to those discussed in the above references.

\section{Lens Parameter Method}

The commonly observed parameters for gravitational lenses are the lens
redshift ($\zl$), the source redshift ($\zs$), the mean deflection of
the lensed object (or similarly the critical radius ($\theta_{crit}$)),
the lens magnitude ($\ml$), and the source magnitude ($m_{S}$).  For
some systems, one or two of these observational parameters may be
missing.

How sensitive is $\theta_{crit}$ to $\Omega_{m}$ and $\Lambda$ for a
given set of $\zl$, $\zs$, and $\ml$?
To calculate $\theta_{crit}$, we will adopt the singular
isothermal sphere (SIS) model for the lens, along with the filled-beam
approximation (see section 4), and the Faber-Jackson relation (Faber \& Jackson
1976). The Faber-Jackson relation relates the velocity dispersion ($\sigma$) of
E/S0 galaxies to their luminosities (L): $\sigma = \sigma_{*}
(L/L_{*})^{\beta}$. Here, we adopt $\sigma_{*}=225 km/sec$, $\beta=0.25$
and $M_{*B_{T}}=-19.9+5log(h)$, taken from Kochanek (1992,1995).

Then $\thc (\zl, \zs, \ml)$ can be expressed as;
\begin{equation}
  \thc = 4\pi (\sigma_{*}/c)^{2}\, d(\zs,\zl)
(d(0,\zl)(1+\zl)^{2})^{4\beta}
/d(0,\zs)\, 10^{-0.8\beta(m-M_{*}-K(z)-E(z)-25)}
\end{equation}
where $d(z_{1},z_{2})$ is the angular size distance between the
redshifts $z_1$ and $z_2$ in Mpc, m is the total apparent magnitude of
the lens galaxy, K(z) and E(z) are the K-correction and the evolutionary
correction, respectively, for the lens galaxy.
The (E+K) correction is important, and to calculate it
we will use the 1 Gyr burst model of Bruzual \& Charlot
(1993) at the formation redshift $z_{for}=10$.
This (E+K) correction is consistent with the results from
the HST Medium Deep Survey (MDS) on the evolution of the luminosity function of
elliptical galaxies (Im et al. 1996), which shows a brightening in
luminosity by about
1 magnitude looking back to $z \sim 1$.

\begin{figure}[bth]
\centerline{\psfig{figure=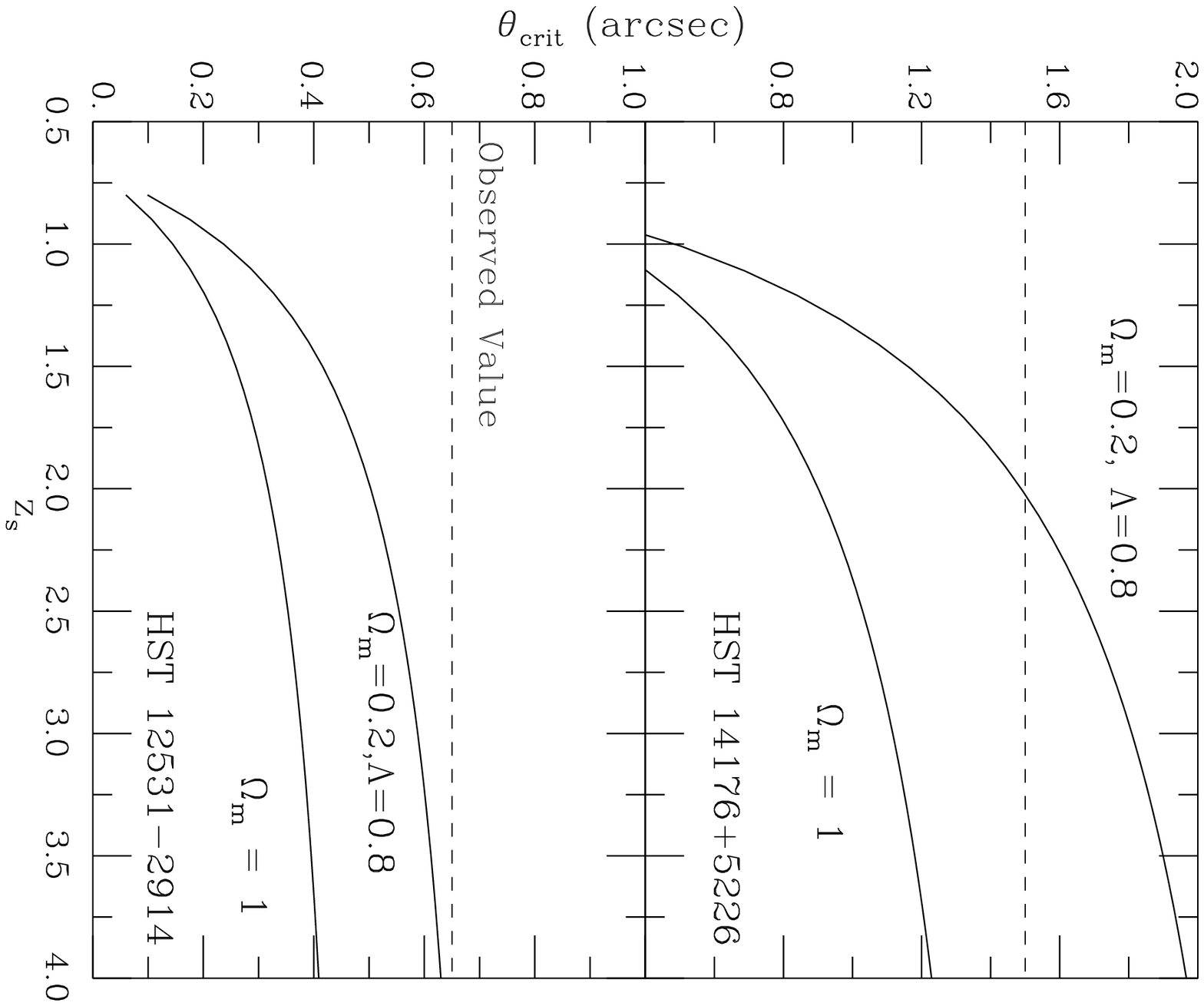,width=4.in,angle=90}}
\footnotesize{
Figure 1: The $\thc$-$\zs$ relation for two gravitational lenses
HST14176+5226 and HST 12531-2914, assuming two different sets of
cosmological parameters, i.e., ($\Omega_{m}=1, \Lambda=0$) and
($\Omega_{m}=0.2, \Lambda=0.8$).  Observed values for $\thc$ are
indicated with the dashed lines.}
\end{figure}

Fig. 1 shows the $\theta_{crit} - \zs$ relation for the strong
gravitational lens systems HST 12531-2914 and HST14176+5226, taking
parameters from Ratnatunga et al. (1995).  The two curves show the model
predictions under the adoption of different cosmological parameters, and
the horizontal line shows the observed value (the source redshift is
unknown).
The value of $\thc$ is quite sensitive to $\Lambda$ when values are known
for $\zl$, $\zs$, and $\ml$.  However, the uncertainty in the prediction
is about a factor of $10^{0.15}$, which arises mainly from the
uncertainties in the Faber-Jackson relation and in the apparent lens
magnitude.  Hence, a single lens system such as HST 12531-2914 cannot be
used alone to measure $\Lambda$.  In order to set a useful limit on
$\Lambda$ with this method, a sample of at least five lenses is required
(e.g., see Kochanek 1992).

In order to combine the information on cosmological parameters from all
available lenses, we therefore construct a likelihood function which is
the product of the probability of each lens having the observed value of
$\thc$ for the given values of $\zs$, $\zl$, $\ml$ and the cosmological
parameters.
This probability $p_{i}(\theta_{o}; \zl, \zs, \ml, \Omega_m, \Lambda)$
is defined as,
\begin{equation}
 p_{i}(\theta_{o}; \zl, \zs, \ml, \Omega_{m}, \Lambda)~\sim~ \int
 G(\theta_{o},\theta_{crit}(z,\zs,m(z)),\sigma_{\theta}) G(z,\zl,\sigma_{\zl})
 G(m(z),\ml,\sigma_{\ml}) dz
\end{equation}
where $\sigma_{\theta}$ is the dispersion in the predicted
$log_{10}(\theta_{crit})$ due to the uncertainty arising from the
Faber-Jackson relation, together with other minor uncertainties,
$\sigma_{\zl}$ is the uncertainly in $\zl$, and
$\sigma_{\ml}$ is the uncertainty in $\ml$. $G(x,x_0,dx)$ is the Gaussian
function with the mean of $x_0$ and the dispersion of dx.
We adopt $\sigma_{\theta} = 0.14$ (or in terms of magnitude, $\sigma
\simeq 0.7$) which is a combination of the uncertainties arising from
the Faber Jackson relation ($0.65 \times 2 = 0.13$: de Zeeuw \& Franx
1991), $M_{*}$ ($0.3 \times 0.2 = 0.06$: Marzke et al. 1994; Loveday et
al. 1992), and the E+K correction ($0.5 \times 0.2 = 0.1$: Im et
al. 1996). When $\zs$ is not available (HST12531-2914), we also
integrate Eq.(2) over $\zs$, assuming a uniform distribution in redshift
space.

Finally, the likelihood function can be written

\begin{equation}
  L = \prod_{j} p_{j, norm}(\theta_{j}; z_{L,j}, z_{S,j}, m_{L,j})
\end{equation}

where $p_{j, norm}$ is the normalized probability of Eq.(2).

We did not adopt the $(3/2)^{0.5}$ factor (hereafter TOG factor) which
was suggested by TOG84 in order to account for the possible difference
between the velocity dispersion of the underlying dark matter and the
luminous material. Recent studies show that this factor is not necessary
(Kochanek 1993,1994; Breimer \& Sanders 1993; Franx 1993).
Independently, we also checked the necessity of the TOG factor by
considering the mean image splittings (see Section 4).

The advantage of this method over the previous methods is the explicit
use of the lens magnitude and the E+K correction, of which the latter
has been observationally constrained only recently (Im et al. 1996;
Pahre, Djorgovski, \& de Carvalho 1996; Bender, Ziegler, \& Bruzual
1996; Barrientos, Schade, \& Lopez-Cruz 1996).  These measurements
enable us to make a reasonably good estimate of the dynamical properties
of each lens galaxy.  The probability of each individual lens having its
unique configuration can then be calculated based on these individual
properties, so that we do not have to use statistical measurements
(e.g., the luminosity function) which may decrease the dependence on the
cosmological parameters when they are averaged over large numbers of
objects. Although our method is, in principle, not as sensitive to the
value of $\Lambda$ as is the lens number count method (e.g., Fukugita et
al. 1992), the latter method is possibly subject to greater
uncertainties (see section 4).  Our method is slightly more susceptible
to a small change in one of the input parameters, but, in common with the
lens redshift method, we have a smaller number of parameters than the
lens count method. In this respect, our method has an edge over the
latter. In particular, the properties of lens galaxies at high redshift
($z \gtrsim 1.5$) are highly uncertain.  They could be dusty enough that
the result from the lens counts might be biased against the non-zero
$\Lambda$ model (Fukugita \& Peebles 1995).  In contrast, the lens
parameter method uses lensing galaxies which lie at $z \lesssim 1$ (see
section 3), and the method is thus less affected by the unknown
properties of high redshift galaxies.

\section{Sample Selection}

Gravitational lenses are selected using the following criteria.

 1) the strong lensing must be caused by a single galaxy lens.  For
example, we do not include 2016+112 in our sample since there are two
lensing galaxies in this system.  Also, we have excluded lens systems
which are clearly influenced by strong perturbations due to cluster
potentials (e.g., 0957+561, B1422+231).

 2) it must be  known that the lens galaxy is likely to be elliptical.
For example, we do not include B0218+357 in our study since there is
good evidence that the lensing galaxy is a spiral or a late-type galaxy
(Patnaik et al. 1993).

 3) the apparent magnitude and the redshift of the lens galaxy must be
known or estimated to reasonable accuracy.  Accurate values for $\ml$
and $\zl$ are important for estimates of the dynamical properties of the
lens galaxy.

  4) For lens candidates that do not have a measured value for $\zs$,
we select only those which show distinctive features such
as rings or crosses.

We find that there are seven strong gravitational lenses that meet these
selection criteria in the published literature, including objects found
in our HST surveys (Table 1).  B1422+231 is excluded from this list
because of the possible cluster perturbation as well as the ambiguity in
the lens redshift ($\zl = 0.64$ from Hammer et al. 1995 vs. $\zl \simeq
0.4$ from Impey et al.  1996).  For MG0414+0534, there have been
speculations that the source redshift of the system is $z \simeq 1.00$
(Burke 1990; Kochanek 1992), but it was later established to be
$\zs=2.64$ (Lawrence et al. 1995), suggesting that the $z \simeq 1$
measurement pertains to the lens galaxy (Surdej \& Soucail 1994).  We
have analyzed the archived HST observations of this system and find a
preliminary result of $R_{F675W}-I_{F814W} = 1.5 \pm 0.2$ for the lens
galaxy (Ratnatunga et al. 1996), suggesting that $\zl=1.2 \pm 0.4$,
consistent with the previous estimates of $\zl = 1$, and hence we will
adopt $\zl=1.2 \pm 0.4$ for this system.  Finally, we have subtracted a
few tenths of a magnitude from some of the quoted lens magnitudes in the
literature, in order to correct for the total apparent magnitude.  When
the uncertainty in the lens magnitude is not quoted in the relevant
reference, errors of about 0.3 magnitude are assigned to these lens
galaxies.

\section{Results and Discussion}

In Fig. 2, we present the relative likelihood of our measurement against
$\Omega_{m}$ for two cases which are of cosmological interest: i)
$\Omega_{m}+\Lambda=1$ and ii) $\Lambda=0$.  Both likelihood functions
are normalized with the maximum likelihood of case (i), and direct
comparison of cases (i) and (ii) is possible using Fig.2.  When a
flat universe is assumed (case (i)), we find that
$\Lambda=0.64_{-0.26}^{+0.15}$, and we exclude the $\Omega_{m}=1$ model
with 97 \% confidence. Also, a universe with $\Lambda \gtrsim 0.9$ is
excluded at the 95 \% confidence level.
If $\Lambda=0$ is assumed (case (ii)), then $\Omega_{m} \simeq 0$ is
favored.  The difference in the likelihood function between a flat
universe with $\Lambda=0.64$ and an open universe with $0< \Omega < 0.3$
is about 0.5 -- 1.  Hence, a flat universe with non-zero $\Lambda$ is
favored over an open universe at 68 \% -- 82 \% confidence.

\begin{figure}[bth]
\centerline{\psfig{figure=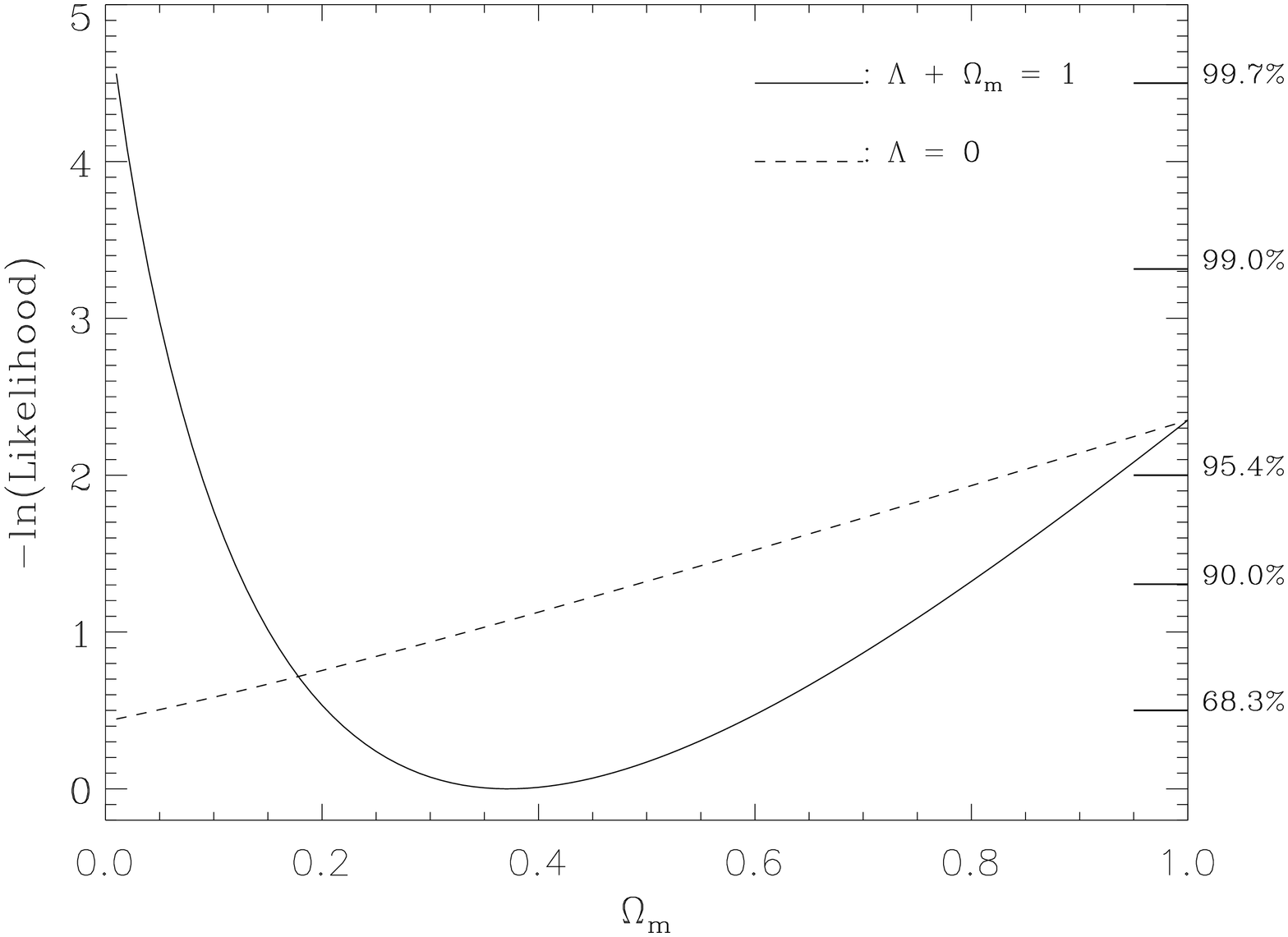,width=4.in,angle=90}}
\footnotesize{
Figure 2: The likelihood of our measurement against $\Omega_{m}$ for two
model universes: i) $\Omega_{m}+\Lambda=1$ (solid line), and ii)
$\Lambda=0$ (dotted line).  The relative likelihoods are normalized with
the maximum of the likelihood of (i).  Confidence limits are indicated
on the right-hand side of the plot, assuming the $\chi^{2}$ distribution
for the likelihood.}
\end{figure}

  Our result is only marginally consistent with the previous estimate of
$\Lambda < 0.7$ based on the lens counts  which strongly favored
 the zero $\Lambda$ flat universe (Kochanek 1995; Maoz \& Rix
1993).    To see what might have caused the disagreement between
our result and the previous results, we have investigated the possible
systematic errors in our analysis and these are listed below:

1) the filled-beam approximation vs. the empty beam approximation

To relate redshift to distance, the filled beam approximation assumes
that light rays propagate through smoothly averaged spacetime.  In
reality, spacetime is inhomogeneous, and therefore the filled beam
approximation may not be correct (e.g., Fukugita et al. 1992).  To see
how our result could be affected by the filled beam approximation, the
analysis was repeated adopting another extreme assumption, namely the
empty beam approximation.  We find that the latter approximation
does not change our result significantly, but strengthens our finding
slightly in favor of the non-zero $\Lambda$ model.

2) Singular isothermal sphere vs. softened isothermal sphere.

We can also assume different mass models for the lens, rather than the
SIS model.  Recent studies show that the SIS model may be too simple to
adequately describe the mass of E/S0s (Lauer 1988; Krauss \& White
1992), although the size of the core radius may be small enough to be
negligible (Wallington \& Narayan 1993).  If the softened isothermal
sphere is used, the predicted $\theta_{crit}$ will be a bit smaller than
the predicted $\thc$ with the SIS model.  In Fig.1, this means that
the predicted lines need to be shifted down along the $\theta_{crit}$ axis,
making the $\Omega_{m}=1$ flat model more inconsistent with the
prediction.  The adoption of the softened isothermal sphere model will
thus strengthen our result.

3) Morphological misclassification

In our analysis, we have assumed that each lens galaxy is an E/S0.  This
assumption may be wrong, and to estimate the bias introduced by treating
a spiral galaxy lens as an elliptical galaxy lens, we repeated our
analysis with the inclusion of one known spiral lens system (B0218+357)
treated as an E/S0 lens.  This caused the result to be strongly biased
in favor of the $\Lambda=0$ model, because of the small predicted $\thc$
of the spiral lens system, a result similar to the issue discussed in
(2).  Thus, if one of the seven lenses we used was actually a spiral
galaxy rather than an elliptical, then the correction of it would only
strengthen our result.

 4) Wrong lens magnitude

Because the lens galaxy is much fainter than the lensed object in some
cases, there is a possibility that the lens magnitudes are not well
determined.  Systematic overestimate of the lens magnitudes by more than
0.6 magnitude can bias the result against the $\Omega_m = 1$ model.
Recent HST observations provide clues as to the accuracy of the
ground-based estimates of lens magnitudes.  The preliminary result by
Falco (1995) from the HST observation of $0142-100$ gives an aperture
magnitude of $R_{F675W}=19.66 \pm 0.01$ for the lens galaxy, while the
original measurement by Surdej et al. (1987) is R=19.  For MG0414+0534,
our preliminary analysis of the HST observation shows $I \simeq 21.4 \pm
0.15$ for the lens galaxy (Ratnatunga et al. 1996), agreeing with the
previous estimate of $I \simeq 21.08 - 21.36$ from the ground (Schechter
\& Moore 1993).  On the other hand, Impey et al. (1996) find $\ml=21.5
\pm 0.3$ in V for the lens galaxy of B1422+231.  The ground based
estimate is $r=21.8 \pm 0.6$ for this object (Yee \& Ellingson 1994; Yee
1995).  At $z \sim 0.4$, $V-r \sim 1$ for E/S0 galaxies, thus the
observed ground-based lens magnitude for this system disagrees with the
HST observation by about 1 magnitude.  These three examples may indicate
that the early ground-based measurements are not very accurate. Errors
seem to go both ways, and hence there may be no systematic overestimate
of the lens magnitudes. But there are only seven lenses in our sample,
and it may be premature to say that there are no systematic errors in
the lens magnitudes.

5) The TOG factor

With the TOG factor included, we find that the peak of the likelihood
function shifts to the region where both $\Omega_{m}$ and $\Lambda$ are
very small, where our test becomes quite insensitive to the cosmological
parameters.  However, many studies have shown in different ways that the
TOG factor is not needed (Kochanek 1995 and refs therein).  In order to
confirm earlier findings, we analyzed the mean image deflections of the
known lens systems with the known source redshifts.  Using criteria (1)
and (2) described in section 3, we find that there are 11 lens systems
available for this analysis (see table 1 in Keeton \& Kochanek 1995).
For these systems, we calculated the ratio $\theta_{crit, obs}/<\thc(\zs)>$.
Since $<\thc (\zs)>$ is fairly independent of the cosmological
parameters (TOG84; Fukugita et al. 1992), the average of 11
$\theta_{crit,obs}/ <\thc(\zs)>$ values will be about 1 if the TOG
factor is not necessary and about 1.5 if the TOG factor is appropriate.
We find an average value of $1.0 \pm 0.1$, confirming that the TOG
factor is not necessary.  In order to test the TOG factor independently,
the study of strong lenses at low redshift ($z < 0.1$) might be
fruitful, since their lens parameters are then insensitive to the
cosmological parameters.  An optical survey which covers a large
fraction of sky (e.g., SDSS) should be able to find a statistically
significant number ($\sim$ 100) of such lenses.

 6) Cluster Perturbation

Since elliptical galaxies preferentially live in a cluster environment,
the gravitational potential of the lens may include a cluster
component. The strong cluster perturbation generally increases the mean
image deflection, and hence we tried to exclude such lenses from our
study (See section 3-(1)).  Nevertheless, we can not completely exclude
the possibility that some of the lenses in our sample include a
considerable amount of cluster perturbation.  If that has happened, our
result could be biased against the $\Omega_{m}=1$ universe.  To understand
the possible contribution to the image splitting from the cluster
potential, detailed modeling of the lens systems is desired using
high resolution images from the HST, or else radio observations.

  7) Source redshift for HST14176+5226

Crampton et al. (1996) have recently published a tentative source
redshift for the lens system HST14176+5226.  A strong emission line is
found at 5324 $\AA$, along with a possible weak emission feature at 6822
$\AA$.  The strong emission line is very likely to be Ly $\alpha$ at
$z=3.39$ if the weak emission feature at 6822 $\AA$ is real, and the latter can
then be identified as CIV 1549.  If the 6822 $\AA$ feature is not real,
then the source object could be located at a redshift lower than
$\zs=3.4$. If $\zs < 3.4$ for the HST14176+5226, then the predicted
$\thc$ will be reduced.  This would bring the peak of the likelihood
function toward the large $\Lambda$ value, strengthening the result in
favor of the non-zero $\Lambda$ model (see Fig. 1) .

   8) E+K correction

  The adopted E+K correction assumes a formation redshift of $z_{for}=10$
 with a 1 Gyr burst of star formation.
  We find that the E+K correction is most sensitive to the value of $z_{for}$,
  and insensitive to the other parameters.
   If we adopt $z_{for} > 10$, then the result changes insignificantly towards
 the zero $\Lambda$ model.
  When $z_{for} < 10$, the result changes in favor of the non-zero $\Lambda$
 model, and the change is significant  when $z_{for} < 2$.
  If $z_{for}=1.5$, $\Lambda$ could be as large as
 $\Lambda \simeq 0.8$.

 If our result is an overestimate in the value of $\Lambda$, then there must
have been a large systematic overestimate of the lens magnitudes and/or
there are strong cluster perturbations.  On the other hand, if the lens counts
have led to an underestimate in the value of $\Lambda$, then that could
have been caused by: i) the dusty nature of high redshift elliptical
galaxies (see section 2 for more discussion), ii) a decrease in the
number density of ellipticals as a function of look-back time, as
expected if most elliptical galaxies were created via major merging
events (Im et al. 1996,1997; Baugh, Cole, \& Frenk 1996; Kauffmann, Charlot,
 \& White 1996), and/or iii)
other uncertainties in the properties of lens galaxies, such as the LF
and the dynamical properties of the low mass ellipticals.  Future HST
observations of faint galaxies, as well as the accumulating redshift data
from ground based telescopes, will hopefully put stringent
constraints on elliptical galaxy evolution at $z > 1$. These data will possibly
give us indications as to why the results from the lens counts have strongly
favored the zero $\Lambda$ model while our result strongly rejects the
flat universe with $\Lambda=0$.  It is noteworthy that neither
method strongly rejects the low $\Omega$ universe.

\section{Conclusions}

We have described and applied the lens parameter method to measure
cosmological parameters using strong gravitational lenses.  Using seven
strong lenses each with an identified lens galaxy, we find that a model
universe with $\Lambda \sim 0.65$ and low $\Omega$ is favored and that
the flat model with $\Lambda=0$ is excluded at greater than 95 \%
confidence.  A universe with low $\Omega$ and $\Lambda=0$ can be
marginally excluded with respect to the flat universe with a non-zero
$\Lambda$ at 68 \% --- 82 \% confidence.  Our result is not biased in
favor of a non-zero $\Lambda$ model due to any conceivable systematic
errors, except for possible strong perturbations from cluster
potentials, and systematic overestimate of the lens magnitudes.  Future
HST observations should uncover new lens systems with measurable lens
properties suitable for this kind of study, and they should also provide
a better understanding of the known lens systems.  We should therefore
be able to get a stronger constraint on $\Lambda$ in the near future.

\acknowledgements

The HST Medium Deep Survey is funded by STScI grants GO2684 {\it et
seqq.}.  We would like to thank the other members of the Medium Deep
Survey Team at JHU, especially Eric J. Ostrander for his efforts on
retrieving and reducing the archival HST data.  We are grateful to
Emilio Falco for providing the lens magnitudes of 0142-100 system. We
also thank Chris Kochanek, Howard K. C. Yee, Joel Primack, and Stefano
Casertano for useful discussions and communications,
 and Mark Subbarao and the anonymous
 referee for helpful comments on the manuscript.

\newpage

\begin{figure}[tb]
\centerline{\psfig{figure=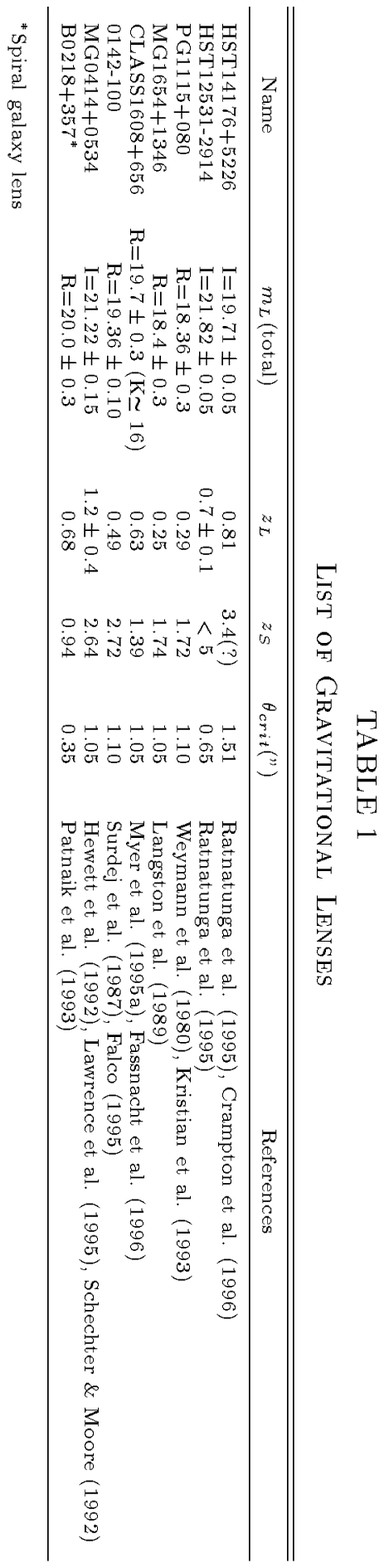}}
\end{figure}

\end{document}